\documentclass[10pt,showpacs,twocolumn]{revtex4}
\usepackage{graphicx}
\usepackage{amsmath}
\usepackage{amssymb}
\usepackage{}
\begin{document}

\preprint{APS/123-QED}


\title{Simultaneous observation of forward-backward attosecond photoelectron holography}

\author{Hongchuan Du$^1$,\footnote{duhch@lzu.edu.cn} Hongmei Wu$^1$, Huiqiao Wang$^2$, Shengjun Yue$^1$,and Bitao Hu$^1$,\footnote{hubt@lzu.edu.cn}
}

\affiliation{%
 $^1$School of Nuclear Science and Technology, Lanzhou University, Lanzhou 730000, China\\
 $^2$Institute of Modern Physics, Chinese Academy of Sciences, Lanzhou 730000, China
}%

\date{\today}

\begin{abstract}
Photoelectron angular momentum distribution of $He^{+}$ driven by a few-cycle laser is investigated numerically. We simultaneously observe two dominant interference patterns with one shot of lasers by solving the 3D time-dependent Schrodinger equation (TDSE). The analysis of a semiclassical model identifies these two interference patterns as two kinds of photoelectron holography. The interference pattern with $P_{z}>0$ is a kind of forward rescattering holography, which comes from the interference between direct (reference) and rescattered (signal) forward electrons ionized in the same quarter-cycle. The interference pattern with $P_{z}<0$ is a kind of backward rescattering holography, which comes from the interference between direct electron ionized in the third quarter-cycle and rescattered backward electron ionized in the first quarter-cycle. Moreover, we propose a method to distinguish this backward rescattering holography and intracycle interference patterns of direct electrons. This is an important step for dynamic imaging of molecular structure and detecting electron in molecules on attosecond time scale by the backward rescattering attosecond photonelectron holography.
\end{abstract}                         
\maketitle


In 1947, Gabor invented the holography [1] in which a coherent beam of light or electron is split into a signal beam and a reference beam. The signal beam scatters off the target and encodes its structure. When it recombines with the reference beam, an interference pattern storing the spatial information of the target is formed. Then the objects can be reconstructed by the interference pattern. Recently, Huismans \emph{et al.} applied the concept of holography to strong-field laser ionization to record temporal and spatial information on the atomic and molecular scale [2], in which a ``forklike" structure is found and recognized as a kind of forward rescattering holography. So far, the backward rescattering holography has not been reported experimentally due to small atomic scattering cross sections. It has been well known that the backward rescattering holography can be used for dynamic imaging of molecular structure [3, 4] and detecting electron motion in molecules or atoms on attosecond time scale [5]. So it is very necessary for observing the backward rescattering holography in experiment. Due to molecular ions with larger rescattering cross sections, Bian $et$ $al.$ theoretically predicted that the backward rescattering holography can be observed by using ${H}_{2}^{+}$[6]. Moreover, they also found that two kinds of forward and backward holography patterns can be simultaneously observed by using the polar molecular ion $HeH^{2+}$ with parallel orientation because of the preferential enhanced ionization from one direction of the molecular axis [7]. Unfortunately, it is still difficult to prepare the target of the molecular ions with perfect orientation in current experimental conditions. Besides, the ground-state $1s\sigma$ of the $HeH^{2+}$ is repulsive and unstable [8, 9]. These limit the use of their scheme. In addition, the interference pattern (as shown in Fig. 5(d)) between the direct electron ionized in the third quarter-cycle and the rescattered backward electron ionized in the first quarter-cycle is similar to the intracycle interference patterns of direct electrons (as shown in Fig. 5(b)). Therefore, the interference pattern asserted as the back holography pattern may come from the intracycle interference patterns of direct electrons. To our best knowledge, how to distinguish these two kinds of interference structures is still an open question.

In this work, we investigate the photoelectron angular momentum distribution of $He^{+}$ driven by a few-cycle laser by solving the TDSE and semiclassical model. It is found that two dominant holography patterns can be observed simultaneously with one shot of lasers. By means of the semiclassical model, we identified the interference pattern with $P_{z}<0$ as a kind of backward rescattering holography, which comes from the interference between the direct electron ionized in the third quarter-cycle and the rescattered backward electron ionized in the first quarter-cycle.

In order to investigate the photoelectron angular momentum distribution, we numerically solve the three-dimensional (3D) time-dependent Schrodinger equation (TDSE) describing the interaction between the linearly polarized laser field and the $He^{+}$ ion. The corresponding TDSE in atomic units (a.u.) is written as
\begin{eqnarray}
i\frac{\partial}{\partial t}\psi(\textbf{r},t)=[-\frac{\nabla^{2}}{2}-\frac{2}{r}+W(\textbf{r},t)]\psi(\textbf{r},t),
\end{eqnarray}
where the interaction term in velocity gauge is defined as $W(r,t)=-i\textbf{A}(t)\cdot\bigtriangledown$. Here, \textbf{A}(t) is the corresponding vector potential of the laser pulse. The time-dependent wave function can be obtained by solving the Eq. (1) in the spherical coordinate system using the Crank-Nicolson method [10]. An absorbing potential is employed to avoid wave function reflection on the grid edge [11, 12].

\begin{figure}[htb]
\centerline{
\includegraphics[width=9.5cm]{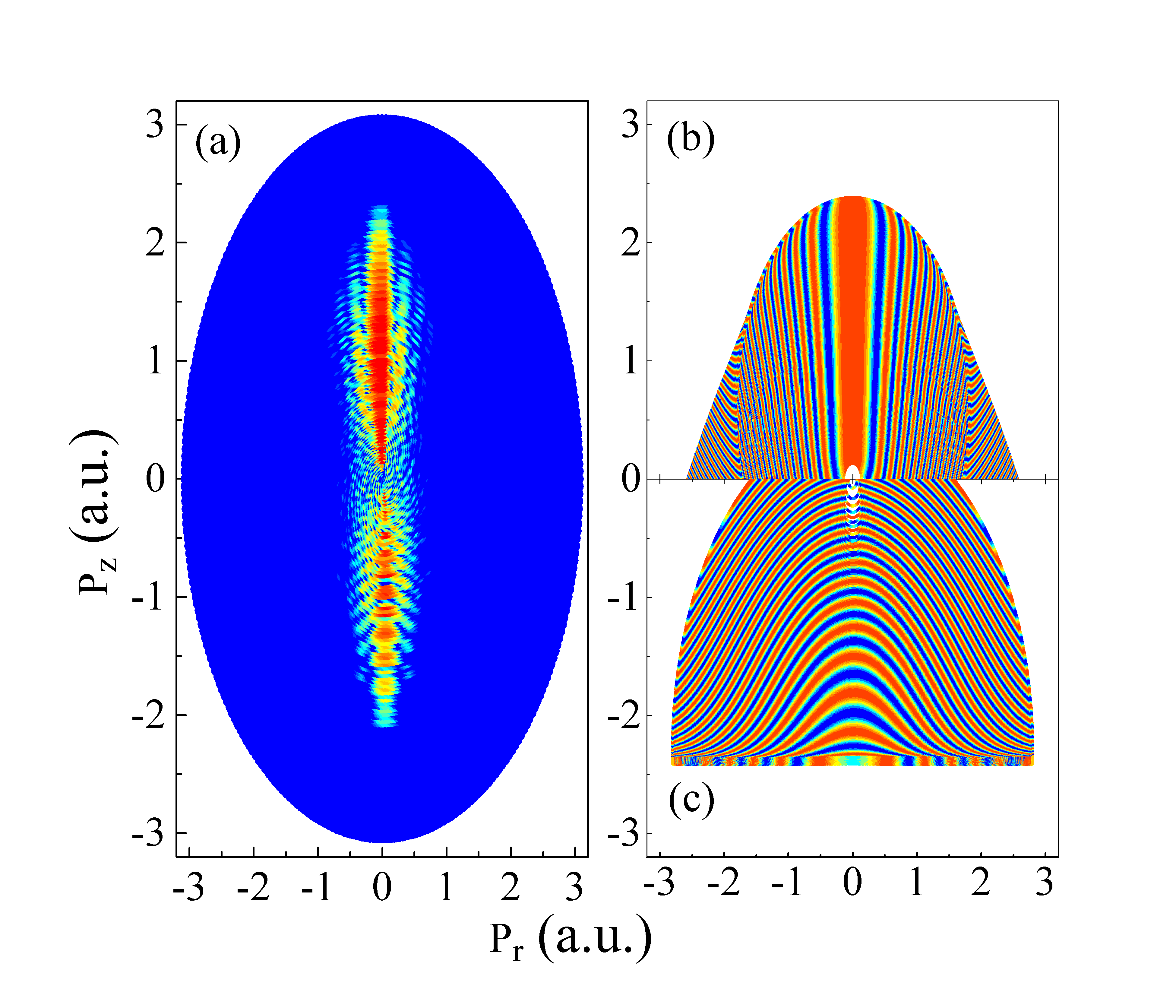}}
 \caption{\label{fig1}(Color online) Photoelectron angular momentum distribution of $He^{+}$ calculated by the TDSE (a) and semiclassical model [(b) and (c)]. (b) represents the interference where the reference electron is generated in the same quarter-cycle, which (c) represents the interference where the reference electron is generated in the third quarter-cycle.
 }
\end{figure}

Since a long-wavelength laser field isn't a necessary condition to attosecond photoelectron holography [14], we adopt a 532-nm laser pulse to reduce the spread of the electron wavefunction in our simulation. The laser intensity is chosen as $1.5\times10^{15}W/cm^{2}$. The vector potential of the laser pulse is given by $A(t)=E_{0}/\omega_{0}\sin^{2}(\pi t/\tau)\sin(\omega_{0}t)$, and the laser polarization direction is along the $z$ direction. $\tau$ is the total duration of the laser pulse, and is set as $5T_{0}$ ($T_{0}$ is the optical cycle of the laser pulse). Figure 1(a) presents the photoelectron angular momentum distribution of $He^{+}$ calculated by projecting the final wavefunction onto plane waves at a time later than the end of the laser pulse. One can clearly see that the photoelectron angular momentum distributions with $P_{z}>0$ and $P_{z}<0$ are quite different since a few-cycle laser pulse is used. In the region of $P_{z}>0$, the interference pattern resembles a ``fork", which is similar to the results in Refs [2, 15]. In their works, the ``fork" structure is identified as a kind of forward rescattering photoelectron holography. However, in the region of $P_{z}<0$, the interference pattern is a semiring structure (downward curvature). Moreover, the width of the stripes and the space between the stripes become larger as the momentum $P_{z}$ decreases from -0.5 to -1.5a.u.. For the sake of checking the accuracy of our results, we also calculate the photoelectron angular momentum distribution by projecting the numerical solution of the TDSE onto the
 \begin{figure}[htb]
\centerline{
\includegraphics[width=7.0cm]{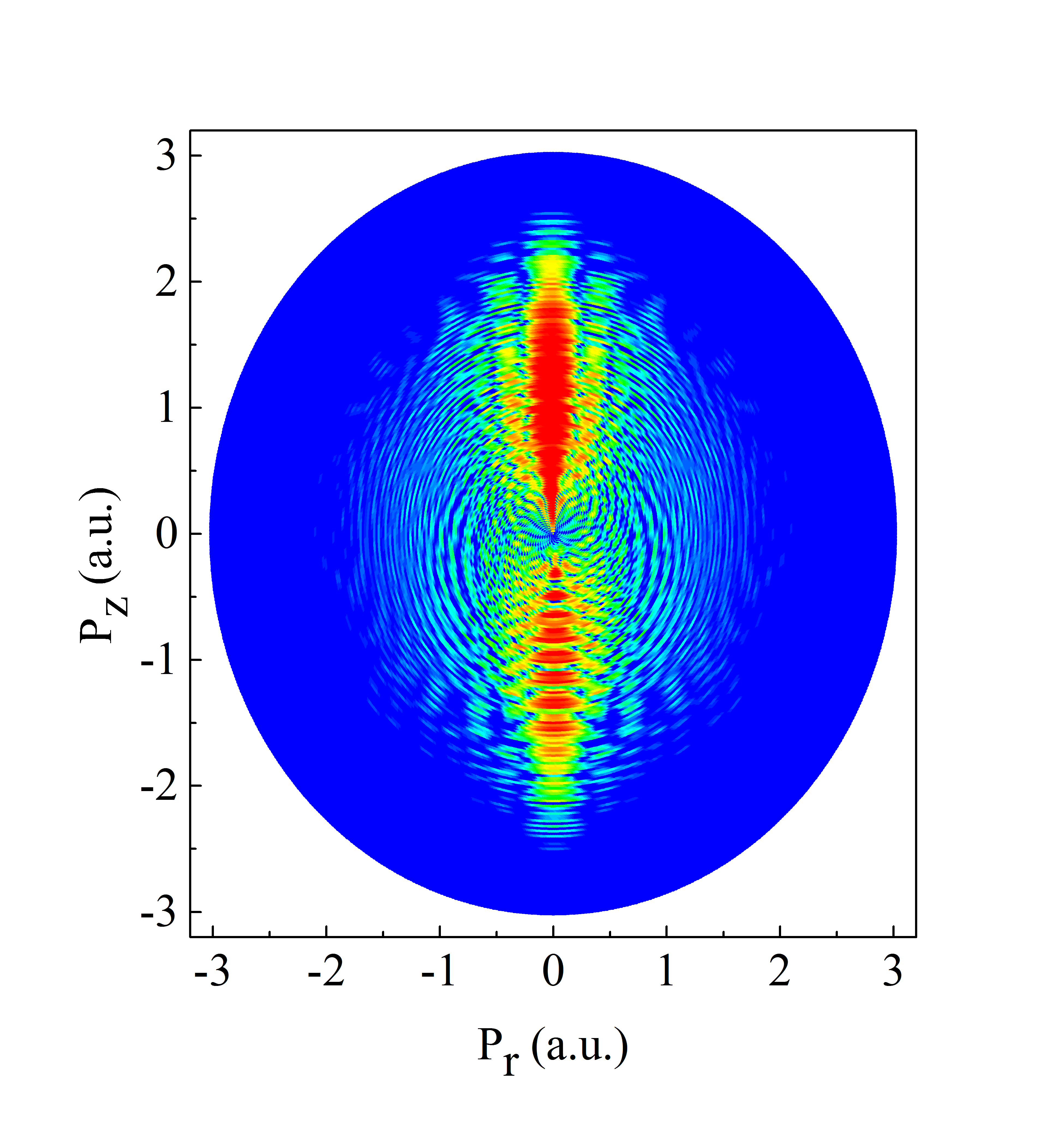}}
 \caption{\label{fig2}(Color online)  Photoelectron angular momentum distribution of $He^{+}$ calculated by projecting the numerical solution of the TDSE onto the incoming Coulomb waves at the end time of the laser pulse.
 }
\end{figure}
incoming Coulomb waves at the end time of the laser pulse, which is shown in Fig. 2. In the calculation, we use a grid with a maximum radius of $r_{max}=2800a.u.$ in the radial direction and maximum number of partial waves $L_{max}=35$. The grid space is 0.02a.u., and the time step is 0.015a.u.. It is found that there is a good agreement between the two methods by comparing with those results in Fig. 1(a). This implies that the results shown in Fig. 1(a) don't come from the calcultaion error. In order to illustrate the physical mechanism of these two kinds of interference patterns, we use the semiclassical recollision model [15, 16], which has been proven to be successful in interpreting forward rescattering photoelectron holography [2, 15]. In the calculation, we used the few-cycle laser pulse. The photoelectron angular momentum distributions calculated by the semiclassical model are presented in Figs. 1(b) and 1(c). Fig. 1(b) shows the interference pattern for the case in Fig. 3(a), where both the signal (A) and reference (B) electrons are generated in the same quarter-cycle. In this case, the signal electron (A) is firstly ionized in the negative z direction, and is then turned around by the laser field. So the velocity of the signal electron (A) is positive before being rescattered. After being
forward rescattered, the signal electron (A) has a positive velocity. Finally, the signal (A) and reference (B) electrons with the same final momentum will interfere with each other and form the interference pattern shown in Fig. 1(b). From this figure, one can clearly see the ``forklike" interference pattern, which is in good agreement with the pattern with $P_{z}>0$ shown in Fig. 1(a). Therefore, the ``forklike" interference pattern with $P_{z}>0$ in Fig. 1(a) comes from the interference between direct (reference) and rescattered (signal) forward photoelectrons ionized in the same quarter-cycle as shown in Fig. 3(a). Fig. 1(c) presents the interference pattern for the case in Fig. 3(b), where the signal electron (A) is ionized in the first quarter-cycle, and the reference electron (B) is ionized in third quarter-cycle. In this case, the signal electron (A) is firstly ionized in the negative $z$ direction, and is then turned around by the laser field. So the velocity of the signal electron (A) is positive before being rescattered. After being backward rescattered, the signal electron (A) has a negative velocity. Finally, the signal (A) and reference (B) electrons with the same final momentum will interfere with each other and form the interference pattern shown in Fig. 1(c). It can be seen from the Fig. 1(c) that the number of interference stripes with $P_{z}\in[-1.5,-0.5]a.u.$ predicted by the semiclassical model is 7. Moreover, the width of the stripes and the space between the stripes calculated by the semiclassical model become larger as the momentum $P_{z}$ decreases from $-0.5$ to $-1.5a.u.$. These agree well with the TDSE results in Fig. 1(a). However, we can't ensure that the interference pattern with $P_{z}<0$ in Fig. 1(a) is caused by the interference in Fig. 1(c) because the intracycle interference pattern of direct electrons has also a similar structure [17, 18]. So far, it is still an open question to distinguish these two kinds of structures.

\begin{figure}[htb]
\centerline{
\includegraphics[width=9.5cm]{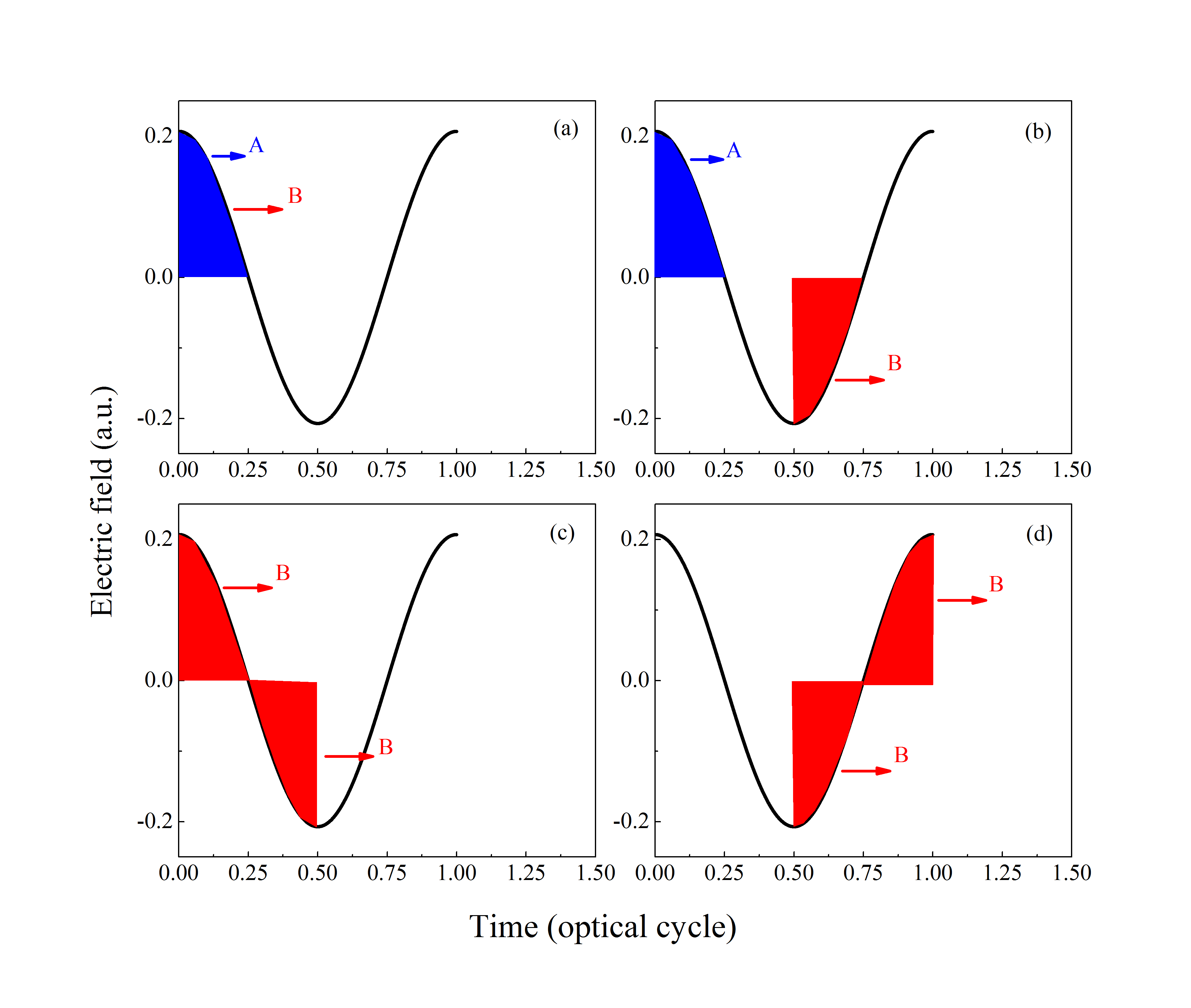}}
 \caption{\label{fig2}(Color online)  Schematic illustration of interference trajectories. A is the rescattering channel, and B is the direct ionization channel.
 }
\end{figure}

As well known, the rescattering process is closely related to the Coulomb potential of parent ion, so the interference patterns will be significantly changed when a short-range potential is used if they are caused by the rescattering process. However, for the interference of direct electrons, the interference pattern is less sensitive to the Coulomb potential of parent ion. Thus we investigate the Coulomb-tail effects to the photoelectron angular momentum distribution in order to confirm that the interference patterns with $P_{z}<0$ in Fig. 1(a) come from the interference in Fig. 3(b). In the calculation, the potential function V(r) is instead with a short-range potential $V_{s}(r)$. We choose
\begin{eqnarray}
V_{s}(r)=-(b+\frac{2.0}{r})exp(-r^{2}/s^{2}),
\end{eqnarray}
where $s=1.6$ and $b=0.57$ to keep the ground-state energy unchanged.

\begin{figure}[htb]
\centerline{
\includegraphics[width=9.5cm]{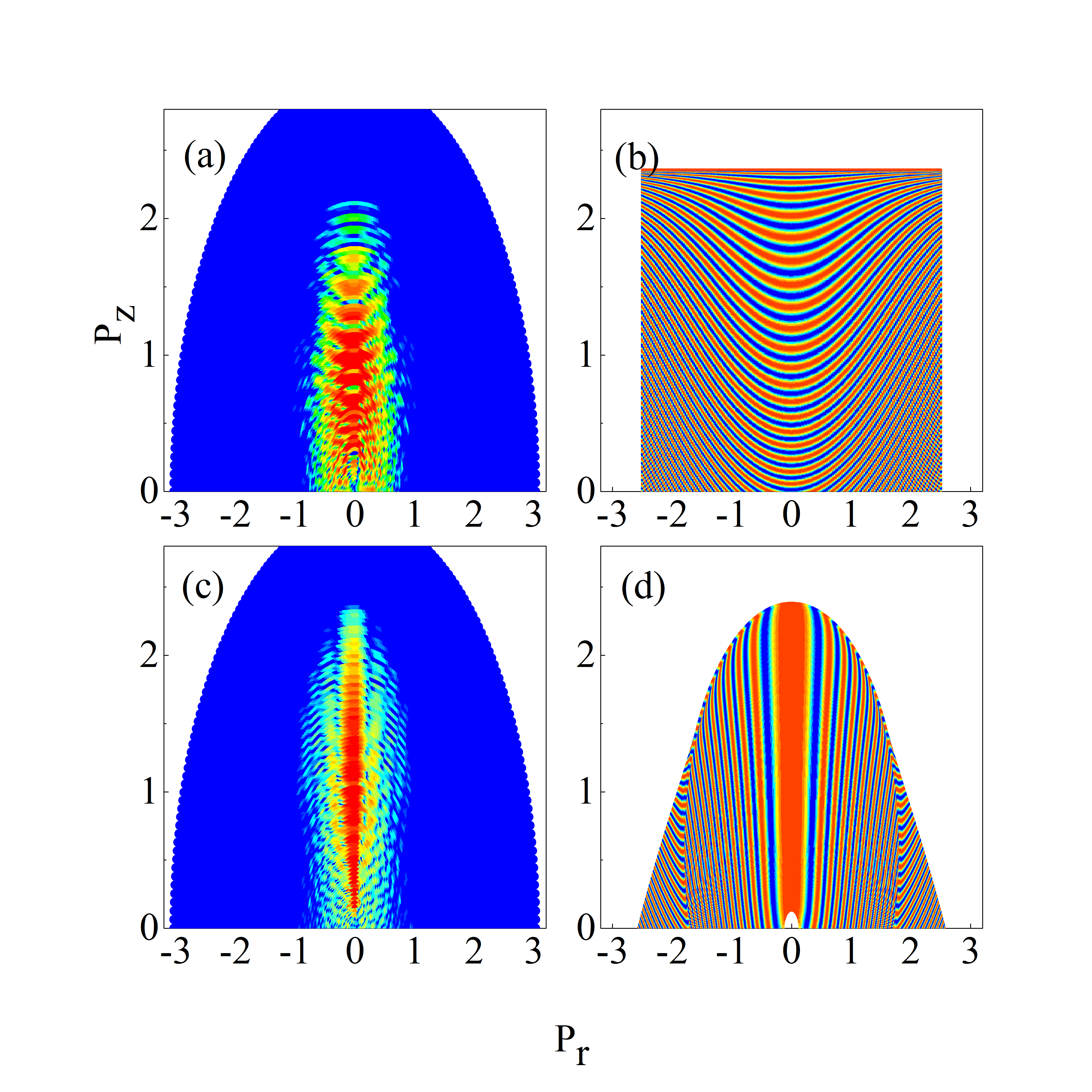}}
 \caption{\label{fig3} (Color online)  Photoelectron angular momentum distribution with $P_{z}>0$ of $He^{+}$ calculated by the (a) TDSE with a short-range potential $V_{s}(r)$, (c) the full TDSE, and semiclassical model [(b) and (d)]. (b) represents the interference between direct electrons generated in first and second quarter-cycle. (d) represents the interference where the reference electron is generated in the same quarter-cycle.
 }
\end{figure}

In order to confirm above conclusion, we firstly compare the photoelectron angular momentum distribution with $P_{z}>0$ of $He^{+}$ calculated using the following theoretical models: (a) solving the TDSE with a short-range potential, (b) using a semiclassical model which describes the interference between direct electrons generated in first and second quarter-cycle as shown in Fig. 3(c), (c) solving the TDSE with the actual potential, (d) using a semiclassical model which describes the interference where the reference electron is generated in the same quarter-cycle as shown in Fig. 3(a). It is clear that there are obvious differences in the photoelectron angular momentum distributions in Figs. 4(a) and 4(c). In Fig. 4(c), one can see a ``fork" structure, which has been identified as a kind of forward rescattering photoelectron holography as shown in Fig. 4(d). When the short-range potential is used, as shown in Fig. 4(a), the ``fork" structure washes out, and a semiring structure (upward curvature) appears. Fig. 4(b) presents the interference pattern of the direct electrons ionized in first and second quarter-cycle as shown in Fig. 3(c). It is seen that the number of interference stripes with $P_{z}\in[-1.5,-0.5]a.u.$ predicted by the semiclassical model is 8, which is in good agreement with those in Fig. 4(a). Hence, we conclude that the semiring structure (upward curvature) in Fig. 4(a) comes from the intracycle inference of direct electrons in Fig. 3(c).

\begin{figure}[htb]
\centerline{
\includegraphics[width=9.5cm]{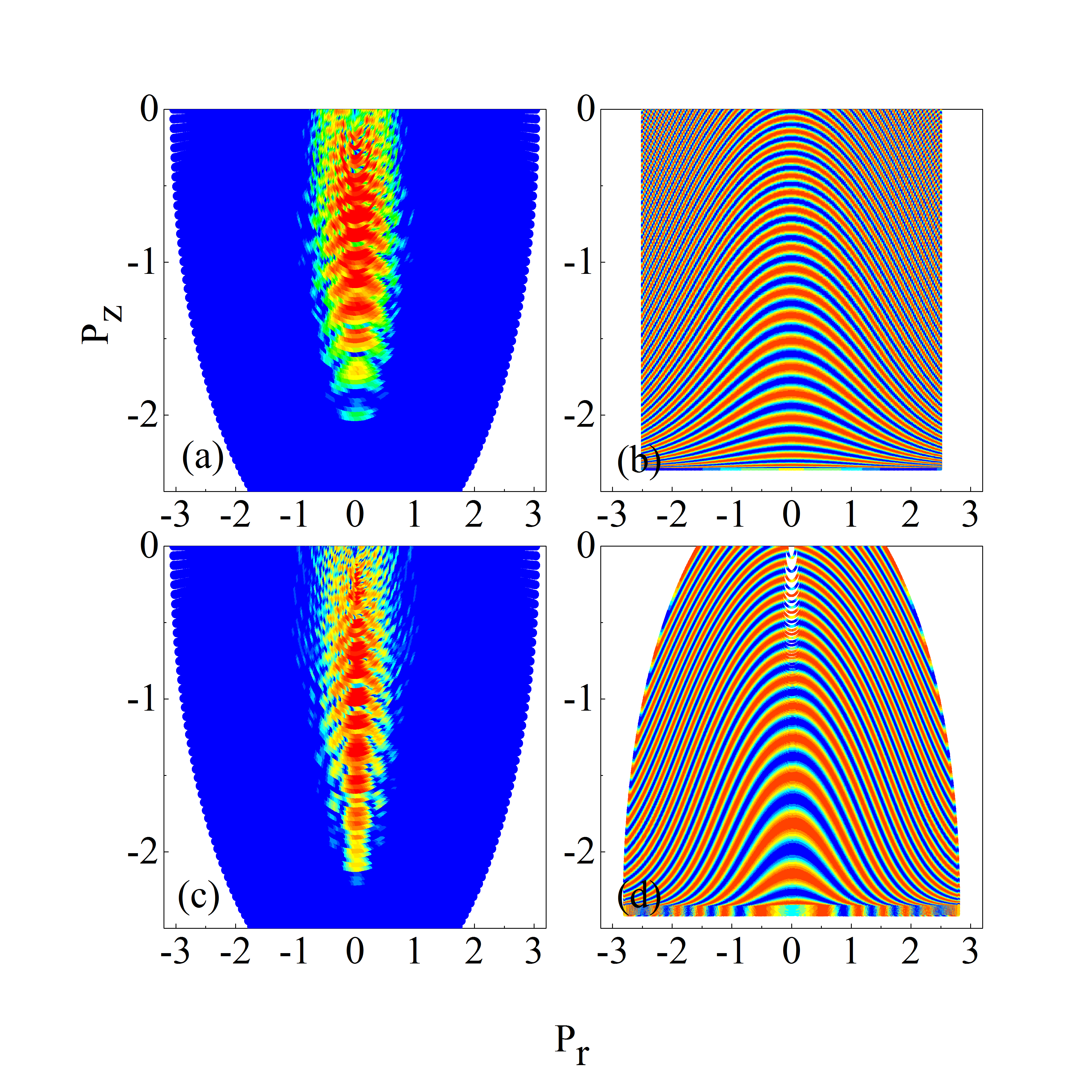}}
 \caption{\label{fig4}(Color online) Photoelectron angular momentum distribution with $P_{z}<0$ of $He^{+}$ calculated by the (a) TDSE with with a short-range potential $V_{s}(r)$, (c) the full TDSE, and semiclassical model [(b) and (d)]. (b) represents the interference between direct electrons generated in third and fourth quarter-cycle. (d) represents the case where the reference electron is generated in the third quarter-cycle.
 }
\end{figure}

Next, we investigate the Coulomb-tail effects to the interference pattern with $P_{z}<0$ in Fig. 1(a). Likewise, we compare the photoelectron angular momentum distribution with $P_{z}<0$ of $He^{+}$ calculated using the following theoretical models: (a) solving the TDSE with a short-range potential, (b) using a semiclassical model which describes the interference between direct electrons generated in third and fourth quarter-cycle as shown in Fig. 3(d), (c) solving the TDSE with the actual potential, (d) using a semiclassical model which describes the interference where the reference electron is generated in the third quarter-cycle as shown in Fig. 3(b). At first glance, the photoelectron angular momentum distributions in Figs. 5(a) and 5(c) appear to be fairly similar. Both of them show a semiring structure (downward curvature). Moreover, the width of the stripes and the space between the stripes become larger as the momentum $P_{z}$ decreases from $-0.5$ to $-1.5a.u.$. However, a closer look reveals that the number and location of stripes are quite different in the two pictures. In Fig. 5(a), the number of interference stripes with $P_{z}\in[-1.5,-0.5]a.u.$ is 8, which is in good agreement with the results predicted by the semiclassical model in Fig. 5(b), in which the interference pattern is produced by the direct electrons ionized in third and fourth quarter-cycle as shown in Fig. 3(d). However, in Fig. 5(c), the number of interference stripes with $P_{z}\in[-1.5,-0.5]a.u.$ is 7 which is in good agreement with the results predicted by semiclassical model in Fig. 5(d), which describes the interference where the reference electron is generated in the third quarter-cycle as shown in Fig. 3(b). Moreover, it is also noted that the curvature of the semiring structure in Fig. 5(c) is larger than that in Fig. 5(a), which is also observed in Figs. 5(b) and 5(d). In addition, it is worth mentioning that the intracycle interference patterns of direct electrons can be shifted by the Coulomb tail. Fortunately, the number of the interference stripes becomes sparse when the Coulomb tail is cut off as shown in Fig. 5 of Ref. [18], which is opposite with our results. Therefore, we conclude that the semiring structure (downward curvature) in Fig. 1(a) comes from the interference between direct electron ionized in the third quarter-cycle and rescattered backward electron ionized in the first quarter-cycle as shown in Fig. 3(b).

In summary, we propose a way of simultaneously observing two kinds of photoelectron holography. The forward rescattering holography comes from the interference between direct and rescattered forward photoelectrons ionized in the same quarter-cycle, which appears in the region of $P_{z}>0$. The backward rescattering holography comes from the interference between direct photoelectron ionized in the third quarter-cycle and rescattered backward photoelectron ionized in the first quarter-cycle, which appears in the region of $P_{z}<0$. Moreover, we also propose a method to distinguish the backward rescatterring holography and the intracycle interference pattern of direct electrons. The laser parameters in this work are experimentally accessible currently. It is predicted that the backward rescattering photoelectron holography can be observed in experiment by in the near future. Finally, we would like to remind that this kind of backward rescattering holography pattern must be carefully confirmed before it is used to dynamic imaging of molecular structure and detecting electron in molecules on attosecond time scale.

\section*{Acknowledgement}

This work was supported by the National Natural Science Foundation of China (Grants Nos. 11404153, No. 11135002, No. 11175076, No. 11475076, and No. 11405077).


\begin{thebibliography}{99}

\bibitem{1} D. Gabor, Nature (London) {\bf161}, 777 (1948).

\bibitem{2} Y. Huismans A. Rouz$\acute{e}$e, A. Gijsbertsen, J. H. Jungmann, A. S. Smolkowska, P. S. W. M. Logman, F. L$\acute{e}$pine, C. Cauchy, S. Zamith, T. Marchenko, J. M. Bakker, G. Berden, B. Redlich, A. F. G. van der Meer, H. G. Muller, W. Vermin, K. J. Schafer, M. Spanner, M. Yu. Ivanov, O. Smirnova, D. Bauer, S. V. Popruzhenko, M. J. J. Vrakking, Science {\bf331}, 61 (2010).

\bibitem{3} T. Zuo, A. D. Bandrauk, and P. B. Corkum, Chem. Phys. Lett. {\bf259}, 313 (1996).

\bibitem{4} A. D. Bandrauk and M. Ivanov, Quantum Dynamic Imaging (Springer, New York, 2011).

\bibitem{5} H. C. Shao and A. F. Starace, Phys. Rev. Lett. {\bf105}, 263201 (2010).

\bibitem{6} X.-B. Bian and A. D. Bandrauk, Phys. Rev. Lett. {\bf108}, 263003 (2012).

\bibitem{7} X.-B. Bian and A. D. Bandrauk, Phys. Rev. A {\bf89}, 033423 (2014).

\bibitem{8} G. Lagmago Kamta and A. D. Bandrauk, Phys. Rev. A 76, 053409 (2007)

\bibitem{9} I. Ben-Itzhak, I. Gertner, O. Heber, and B. Rosner, Phys. Rev. Lett. {\bf71}, 1347 (1993).

\bibitem{10} J. Crank and P. Nicolson, Proc. Cambridge Philos. Soc. {\bf43}, 50 (1947).

\bibitem{11} T. Gonzale-Lezana, E. J. Rackham, and D. E. Manolopoulos, J. Chem. Phys. {\bf120}, 2247 (2004).

\bibitem{12} D. E. Manolopoulos, J. Chem. Phys. {\bf117}, 9552 (2002).

\bibitem{13} T. Marchenko, Y. Huismans, K. J. Schafer, and M. J. J. Vrakking, Phys. Rev. A {\bf84}, 053427 (2011).

\bibitem{14} X.-B. Bian, Y. Huismans, O. Smirnova, K.-J Yuan, M. J. J. Vrakking, and A. D. Bandrauk, Phys. Rev. A {\bf84}, 043420 (2011).


\bibitem{15}P. B. Corkum, Phys. Rev. Lett. {\bf71}, 1994 (1993).

\bibitem{16} X. Xie, S. Roither, D. Kartashov, E. Persson, D. G. Arbo, L. Zhang, S. Grafe, M. S. Schoffler, J. Burgdorfer, A. Baltuska, and M. Kitzler, Phys. Rev. Lett. {\bf108}, 193004 (2012).

\bibitem{17} D. G. Arbo,  K. L. Ishikawa,  K. Schiessl,  E. Persson, and J. Burgdorfer, Phys. Rev. A {\bf81}, 021403 (2010).


\end{thebibliography}
\end{document}